# Power Series Solution to Non-Linear Partial Differential Equations of Mathematical Physics

[a]E. López-Sandoval[*a], A. Mello[a], J. J. Godina- Nava[b]
[a] Centro Brasileiro de Pesquisas Físicas,
Rua Dr. Xavier Sigaud, 150
CEP 22290-180, Rio de Janeiro, RJ, Brazil.
*sandoval@cbpf.br
[b]Departamento de Física
Centro de Investigación y de Estudios Avanzados
del Instituto Politecnico Nacional
Ap. Postal 14-740, 07000, México, D. F. México

**Abstract**

Power Series Solution method has been traditionally used to solve Linear Differential Equations: in Ordinary and Partial form. However, despite their usefulness the application of this method has been limited to this particular kind of equations. We propose to use the method of power series to solve non-linear partial differential equations. We apply the method in several typical non linear partial differential equations in order to demonstrate the power of the method.

# 1 Introduction

Nowadays, the solution of non-linear partial differential equations is considered as a fundamental tool in the research of multidisciplinary areas, because both their implication in the public health problems and social impact in to solve real life problems. In fact, is mandatory to involve mathematical methods in the traditional research methodology of science areas like Biology, Cell Biology, Physiology, Physics, Chemistry, Chemical Physics, etc. which helped by the technological advance in the computation, to incorporate a new age of knowledge in order to tackle real problems.

Power Series Solution (PSS) method (PSSM) has been limited to solve Linear Differential equations, both Ordinary (ODE) [1, 2], and Partial (PDE) [3, 4]. Linear PDE has traditionally been solved using the variable separation method because it permits to obtain a coupled system of ODE easier to solve with the PSSM. Examples of these are the Legendre polynomials and the spherical harmonics used in the Laplace´s Equations in spherical coordinates or the Bessel´s equations in cylindrical coordinates [3-4]. It is known that in Non Linear PDE (NLPDE) is, as we know, because isn´t possible to apply the separation of variables method.

The methodology to solve the NLPDE is to obtain a solution by using the approximated analytical method, *i. e.,* non numerical or semi analytic form, in a indirect, or direct way. In the direct way, there are methods like Inverse Scattering Transform [5] or the Lax Operator Formalism [6]. In the direct way it can be used for example the PSS in an asymptotic approximation the Hirota method involving a bilinear operator technique [7]; the Adomian Decomposition Method [8], and the Homotopy Analysis method [9, 10]. This last method involves a series expansion with with a non small parameter perturbation approximation to adjust the convergence. This method is different of the classical perturbation theory.

Techniques, even more direct, to approximate to a solution in NLPDE are the Taylor Polynomial Approximation method (TPAM) [11, 12], and the PSSM. In both techniques a semi analytic solution is obtained implementing the PSSM. However, the PSSM has been little used to solve non-linear ODE [13-16], or NLPDE [17-19].

In this letter, we propose to apply in order to find particular solutions of typical PDE widely used in mathematical physics, namely, the equation of a steady state laminar boundary layer on a flat plate, the Burger's equation and the Korteveg-de Vries equations. In all this examples, we were able to find particular values of the coefficients of the truncated PSSM. We use the symbolic computation package Matlab® to obtaining the algebraic operations for the truncated series approximation. This program helps to do easier the tedious algebraic operations.

## 2 The Power Series Solution method

We know that almost the totality of the NLPDE have not a solution with an analytic expression, *i.e.* a solution in terms of know functions in a closed form. Our proposal is to construct one solution using a power series, taking advantage of the capacity of power series to represent any function with an algebraic series and develop the idea to construct an approximate solution. It also has the possibility to approximate a solution, inclusive if it not exists in analytic form. In a similar way in which we apply normally the Taylor series (TS) to some function:

$$f(x_1,...,x_d) = \sum_{n_1=0}^{\infty}\sum_{n_2=0}^{\infty}...\sum_{n_d}^{\infty} a_{n_1..n_d}(x_1-a_1)^{n_1}...(x_d-a_d)^{n_d} \qquad (1)$$

where

$$a_{n_1..n_d} = \frac{\left(\frac{\partial^{n_1+...+n_d} f}{\partial x_1^{n_1}...\partial x_d^{n_d}}\right)(a_1,...,a_d)}{n_1!n_2!...n_d!} \qquad (2)$$

are their corresponding coefficient´s expansions. It is necessary to know the values of all the derivatives of the function at $(a_1,...,a_d)$ that represent its center in a open disc. The Taylor´s series not always guarantee per se that the function represented has an exact approximation to distant points about its central value. However, considering that the PSS need to satisfy the NLPDE with an initial values condition (IVC) in time or space values, or with the boundary values conditions (BVC) in the space, we can to construct a well posed problem for to obtain an accurate solution. Also the Taylor´s series approximation is a smoothed function and therefore, with this we can to guarantee the existence of a solution.

When the differential equation (DE) under study is an IVC problem, it is possible to obtain a solution finding these IVC (derivatives of the first orders as is showed in the eq. (2)), using the DE to obtain the higher derivatives orders, and substitute them in eq. (1); but this procedure is useful only to solve trivial problems [1]. That disadvantage can be overcome using the PSSM, because this method make possible to obtain the others values with the recurrence relation technique.

The PSSM represent a general solution with unknown coefficients, and when the equation (1) is substituted in the PDE we obtain a recurrence relation for the expansion coefficients. These coefficients in their form should be expressed in function of the coefficients result of the IVC application (because as we say before, the coefficients of the TS are the derivative of the function solution, Eq. (1)). But these coefficients also can be obtained with the evaluation of the BVC, or of any other point in the space whose value are known, and it will depend on the kind of the problem under study. In this way, we obtain a system of equations in function of these initial value based coefficients. The number of equations should be correlated with the data, i. e., with the IVC or BVC. In order to obtain and to solve a coherent algebraic system of equations, we also need the same number of both: coefficients and equation. All these conditions in first instance help us to guarantee that the PDE is a well posed problem, *i. e.*, the existence, uniqueness and smoothness of its solution is well defined [20].

The PSSM is a proposal to find a semi analytic solution as an asymptotic approximation (in the space and time) of a finite series with a minimal error in the expansion of terms of the series.

## 3 The PSSM applied to NLPDE

In order to illustrate the use of the PSSM, we start exemplifying with the solution of the equation of a steady state laminar boundary layer on a flat plate [21, 22]:

$$\frac{\partial U}{\partial y}\frac{\partial^2 U}{\partial yx} - \frac{\partial U}{\partial x}\frac{\partial^2 U}{\partial y^2} = \nu \frac{\partial^3 U}{\partial y^3} \qquad (3)$$

The PDE has two variables *x* and *y*, therefore the proposal for PSS is:

$$U(x,y) = \sum_n \sum_m a_{nm} x^n y^m \qquad (4)$$

Substituting eq. (4) in (3):

$$\sum_n \sum_m m a_{nm} x^n y^{m-1} \sum_n \sum_m m a_{nm} x^{n-1} y^{m-1} - \sum_n \sum_m n a_{nm} x^{n-1} y^m \sum_n \sum_m m(m-1) a_{nm} x^n y^{m-2} = \\ \nu \sum_n \sum_m m(m-1)(m-2) a_{nm} x^n y^{m-3} \qquad (5)$$

We use the symbolic algebraic solver of Matlab® to solve the series multiplication until reach the *l=4* (*l=n+m=4*) power degree, following the method. Solving for each of the terms and matching each coefficient of the same power degree, we obtain:

l=0
$$a_{01}a_{11} - 2a_{02}a_{10} = 6\nu a_{03}$$
l=1
$$a_{11}^2 + 2a_{01}a_{21} - 4a_{02}a_{20} - 2a_{10}a_{12} = 6a_{13}\nu$$

$$2a_{01}a_{12} - 6a_{03}a_{10} = 24\nu a_{04}$$

$$l=2$$

$$4a_{01}a_{22} - 12a_{03}a_{20} - 6a_{10}a_{13} + 2a_{11}a_{12} = 24a_{14}\nu$$

$$-6a_{23}\nu = 0$$

$$3a_{01}a_{13} + 2a_{02}a_{12} - 3a_{03}a_{11} - 12a_{04}a_{10} = 60a_{05}\nu$$

$$l=3 \tag{13}$$

$$4a_{01}a_{14} + 4a_{02}a_{13} - 8a_{04}a_{11} - 20a_{05}a_{10} = 120a_{06}\nu$$

$$2a_{12}^{2} + 6a_{01}a_{23} + 4a_{02}a_{22} - 6a_{03}a_{21} - 24a_{04}a_{20} - 12a_{10}a_{14} = 60a_{15}\nu$$

$$6a_{01}a_{32} - 18a_{03}a_{30} - 6a_{10}a_{23} + 4a_{11}a_{22} + 2a_{12}a_{21} - 12a_{13}a_{20} = 24a_{24}\nu$$

$$2a_{21}^{2} + 4a_{01}a_{41} - 8a_{02}a_{40} - 2a_{10}a_{32} + 4a_{11}a_{31} - 6a_{12}a_{30} - 4a_{20}a_{22} = 6a_{33}\nu$$

$$l=4$$

$$5a_{01}a_{15} + 6a_{02}a_{14} + 3a_{03}a_{13} - 4a_{04}a_{12} - 15a_{05}a_{11} - 30a_{06}a_{10} = 210a_{07}\nu$$

$$8a_{01}a_{42} - 24a_{03}a_{40} - 6a_{10}a_{33} + 6a_{11}a_{32} + 2a_{12}a_{31} - 18a_{13}a_{30} - 12a_{20}a_{23} + 4a_{21}a_{22} = 24a_{34}\nu$$

$$9a_{01}a_{33} + 6a_{02}a_{32} - 9a_{03}a_{31} - 36a_{04}a_{30} - 12a_{10}a_{24} + 3a_{11}a_{23} + 6a_{12}a_{22} - 3a_{13}a_{21} - 24a_{14}a_{20} = 60a_{25}\nu$$

In result, we have 13 equations with a set of 13 unknown variables ($a_{03}$, $a_{13}$, $a_{04}$, $a_{14}$, $a_{23}$, $a_{05}$, $a_{33}$, $a_{06}$, $a_{15}$, $a_{24}$, $a_{07}$, $a_{34}$, $a_{25}$) one physical parameter ($\nu$), and 8 known variables ($a_{00}$, $a_{01}$, $a_{10}$, $a_{11}$, $a_{02}$, $a_{20}$, $a_{12}$, $a_{21}$). We could know this set of variables applying the IVC or BVC. Solving this system of equations using newly the symbolic algebraic solver of Matlab® to obtain the recurrence relation in function of the known variables and the physical parameter:

$$a_{22} = 0, a_{32} = 0, a_{41} = 0, a_{40} = 0, a_{31} = 0, a_{51} = 0, a_{50} = 0, a_{42} = 0, a_{30} = 0$$

$$a_{03} = \frac{a_{01}a_{11} - 2a_{02}a_{10}}{6\nu}$$

$$a_{13} = \frac{a_{11}^{2} + 2a_{01}a_{21} - 4a_{02}a_{20} - 2a_{10}a_{12}}{6\upsilon}$$

$$a_{04} = \frac{2a_{02}a_{10}^{2} - a_{01}a_{11}a_{10} + 2a_{01}a_{12}\nu}{24\nu^{2}}$$

$$a_{14} = \frac{2a_{10}^{2}a_{12} - a_{10}a_{11}^{2} - 2a_{01}a_{10}a_{21} - 2a_{01}a_{11}a_{20} + 8a_{02}a_{10}a_{20} + 2a_{11}a_{12}\,\nu}{24\nu^{2}}$$

$$a_{05} = \frac{2a_{21}a_{01}^{2}\nu + a_{11}a_{01}a_{10}^{2} - 4a_{12}a_{01}a_{10}\nu - 4a_{02}a_{20}a_{01}\nu - 2a_{02}a_{10}^{3} + 2a_{02}a_{11}a_{10}\nu + 4a_{02}a_{12}\nu^{2}}{120\nu^{3}}$$

$$a_{33} = \frac{a_{21}^{2}}{3\nu}$$

$$a_{34} = \frac{-2a_{10}a_{21}^{2}}{24\nu^{2}}$$

$$a_{24} = \frac{8a_{02}a_{20}^2 - 2a_{11}^2 a_{20} - 4a_{01}a_{20}a_{21} + 4a_{10}a_{12}a_{20} + 2a_{12}a_{21}\nu}{24\nu^2}$$

$$a_{06} = (-4a_{21}a_{01}^2 a_{10}\nu - 2a_{20}a_{01}^2 a_{11}\nu + 12a_{20}a_{01}a_{02}a_{10}\nu + 8a_{21}a_{01}a_{02}\nu^2 - a_{01}a_{10}^3 a_{11} + 6a_{12}a_{01}a_{10}^2 \nu$$
$$+ a_{01}a_{10}a_{11}^2 \nu - 2a_{12}a_{01}a_{11}\nu^2 - 16a_{20}a_{02}^2 \nu^2 + 2a_{02}a_{10}^4 - 6a_{02}a_{10}^2 a_{11}\nu - 12a_{12}a_{02}a_{10}\nu^2 + 4a_{02}a_{11}^2 \nu^2)/720\nu^4$$

$$a_{15} = -(2a_{10}^3 a_{12} - a_{10}^2 a_{11}^2 - 4a_{12}^2 \nu^2 - 2a_{01}a_{10}^2 a_{21} + 12a_{02}a_{10}^2 a_{20} - 4a_{01}a_{10}a_{11}a_{20} + 2a_{01}a_{11}a_{21}\nu$$
$$+ 4a_{01}a_{12}a_{20}\nu - 4a_{02}a_{10}a_{21}\nu + 2a_{10}a_{11}a_{12}\nu)/(120\nu^3)$$

$$a_{25} = (4a_{01}a_{11}a_{20}^2 - 24a_{02}a_{10}a_{20}^2 + 4a_{10}a_{11}^2 a_{20} - 8a_{10}^2 a_{12}a_{20} + 4a_{01}a_{21}^2 \nu - a_{11}^2 a_{21}\nu + 8a_{01}a_{10}a_{20}a_{21}$$
$$+ 4a_{02}a_{20}a_{21}\nu + 4a_{11}a_{12}a_{20}\nu)/(120\nu^3)$$

$$a_{07} = -(-6a_{21}a_{01}^2 a_{10}^2 \nu - 6a_{20}a_{01}^2 a_{10}a_{11}\nu + 4a_{21}a_{01}^2 a_{11}\nu^2 + 4a_{20}a_{01}^2 a_{12}\nu^2 + 24a_{20}a_{01}a_{02}a_{10}^2 \nu +$$
$$24a_{21}a_{01}a_{02}a_{10}\nu^2 + 8a_{20}a_{01}a_{02}a_{11}\nu^2 - a_{01}a_{10}^4 a_{11} + 8a_{01}a_{10}^3 a_{12}\nu + 3a_{01}a_{10}^2 a_{11}^2 \nu - 12a_{01}a_{10}a_{11}a_{12}\nu^2 -$$
$$2a_{01}a_{11}^3 \nu^2 + 4a_{01}a_{12}^2 \nu^3 - 80a_{20}a_{02}^2 a_{10}\nu^2 + 2a_{02}a_{10}^5 - 12a_{02}a_{10}^3 a_{11}\nu - 24a_{02}a_{10}^2 a_{12}\nu^2$$
$$+ 20a_{02}a_{10}a_{11}^2 \nu^2)/(5040\nu^5)$$

...

Therefore, our result is:

$$U(x, y) = a_{00} + a_{10}x + a_{01}y + a_{11}xy + a_{20}x^2 + a_{02}y^2 + a_{21}x^2 y + a_{12}xy^2 + a_{03}y^3 + a_{13}xy^3 + a_{04}x^4 + a_{14}xy^4 +$$
$$a_{05}y^5 + a_{24}x^2 y^4 + a_{06}y^6 + a_{15}xy^5 + a_{33}x^3 y^3 + a_{07}y^7 + a_{34}x^3 y^4 + a_{25}x^2 y^5 + ...$$

(8)

The next example is the solution of the Burger equation [23, 24]:

$$\frac{\partial U}{\partial t} + U\frac{\partial U}{\partial x} = \upsilon \frac{\partial^2 U}{\partial x^2}$$

(9)

where $\upsilon$ is the viscosity constant. First, we consider the stationary state of the problem, i. e., the time independent equation:

$$U\frac{\partial U}{\partial x} = \upsilon \frac{\partial^2 U}{\partial x^2}$$

(10)

The series solution proposal is:

$$U(x) = \sum_{n=0}^{\infty} a_n x^n$$

(11)

Substituting Eq. (11) in Eq. (10), we obtain:

$$\sum_{n=0}^{\infty} a_n x^n \sum_{n=1}^{\infty} n a_n x^{n-1} = \sum_{n=1}^{\infty} n(n-1) a_{2n+1} x^{n-2} \quad (12)$$

matching coefficients with the equal exponents, until reach power degree, using Matlab® in a similar way as in the previous example. We obtain the following algebraic equations:

$$\begin{aligned}
&\text{n=0}\\
&a_0 a_1 = 2 v a_2\\
&\text{n=1}\\
&a_1^2 + 2 a_0 a_2 = 6 v a_3\\
&\text{n=2}\\
&a_1 a_2 + a_0 a_3 = 4 v a_4\\
&\text{n=3}\\
&a_1 a_3 + 2 a_2^2 + a_0 a_4 = 5 v a_5\\
&\text{n=4}\\
&a_1 a_4 + a_2 a_3 + a_0 a_5 = 6 v a_6\\
&\text{n=5}\\
&2 a_1 a_5 + 2 a_2 a_4 + a_3^2 + 2 a_0 a_6 = 14 v a_7\\
&\text{n=5}\\
&2 a_1 a_5 + 2 a_2 a_4 + a_3^2 + 2 a_0 a_6 = 14 v a_7\\
&\text{n=6}\\
&a_1 a_6 + a_2 a_5 + a_3 a_4 + a_0 a_7 = 8 v a_8\\
&\text{n=7}\\
&2 a_1 a_7 + 2 a_2 a_6 + 2 a_3 a_5 + a_4^2 + 2 a_0 a_8 = 18 v a_9\\
&\text{n=8}\\
&a_1 a_8 + a_2 a_7 + a_3 a_6 + a_4 a_5 + a_0 a_9 = 10 v a_{10}\\
&\ldots
\end{aligned} \quad (13)$$

Solving the algebraic system of 9 equations, for the 9 unknown variables ($a_2$, $a_3$, $a_4$, $a_5$, $a_6$, $a_7$, $a_8$, $a_9$, $a_{10}$) in function of the known set of coefficients ($a_0$, $a_1$) and the physical parameter $v$, we find that the coefficients of the PSS are:

$$\begin{aligned}
a_2 &= a_0 a_1 / 2v\\
a_3 &= (a_0^2 a_1 + v a_1^2)/6v^2\\
a_4 &= (a_0^3 a_1 + 4 v a_0 a_1^2)/24 v^3
\end{aligned}$$

$$a_5 = (a_0^4 a_1 + 11\nu a_0^2 a_1^2 + 4\nu^2 a_1^3)/120\nu^4 \tag{14}$$

$$a_6 = (a_0^5 a_1 + 26\nu a_0^3 a_1^2 + 34\nu^2 a_0 a_1^3)/720\nu^5$$

$$a_7 = (a_0^6 a_1 + 57\nu a_0^4 a_1^2 + 180\nu^2 a_0^2 a_1^3 + 34\nu^3 a_1^4)/5040\nu^6$$

$$a_8 = (a_0^7 a_1 + 120\nu a_0^5 a_1^2 + 768\nu^2 a_0^3 a_1^3 + 496\nu^3 a_0 a_1^4)/40320\nu^7$$

$$a_9 = (a_0^8 a_1 + 247\nu a_0^6 a_1^2 + 2904\nu^2 a_0^4 a_1^3 + 4288\nu^3 a_0^2 a_1^4 + 496\nu^4 a_1^5)/362880\nu^8$$

$$a_{10} = (a_0^9 a_1 + 502\nu a_0^7 a_1^2 + 10194\nu^2 a_0^5 a_1^3 + 28768\nu^3 a_0^3 a_1^4 + 11056\nu^4 a_0 a_1^5)/3628800\nu^9$$

$$...$$

In consequence, we can approximate freely this solution because the equations have an infinite series solution. Also it is possible to obtain a particular solution considering that $a_0=0$:

$$a_3 = a_1^2/6\nu$$

$$a_5 = a_1^3/30\nu^2$$

$$a_7 = 17 a_1^4/2520\nu^3$$

$$a_9 = 31 a_1^5/22680\nu^4$$

$$\ldots$$

substituting in Eq. (3) we obtain:

$$U(x) = a_1 x + \frac{a_1^2}{6\nu} x^3 + \frac{a_1^3}{30\nu^2} x^5 + \frac{17 a_1^4}{2520\nu^3} x^7 + \frac{31 a_1^5}{22680\nu^4} x^9 + \ldots \tag{15}$$

This particular solution has an odd symmetry own to the system described for the Burger's equation.

Now, we will try to solve the Burger's DE, but considering it dependent of the time, Eq. (9), and then the series solution proposal is:

$$U(x,t) = \sum_n \sum_m a_{nm} x^n t^m \tag{16}$$

Now, substituting Eq. (16) in Eq. (9), we obtain:

$$\sum_n \sum_m m a_{nm} x^n t^{m-1} + \sum_n \sum_m a_{nm} x^n t^m \sum_n \sum_m n a_{nm} x^{n-1} t^m =$$

$$\nu \sum_n 2n(n-1) \sum_m a_{nm} x^{n-2} t^m$$

developing the multiplication of series until reach the values ($n=3$, $m=3$) and matching coefficients with the equals exponents, we obtain:

$$2va_{20}=a_{01}+a_{00}a_{10}$$

$$6va_{30}=a_{10}^2+a_{11}+2a_{00}a_{20}$$

$$2va_{21}= 2a_{02}+ a_{01}a_{10}+ a_{00}a_{11}$$

$$3va_{31}=a_{10}a_{11}+ a_{12}+ a_{01}a_{20}+ a_{00}a_{21}$$

$$- a_{10}a_{20} - a_{21} - a_{00}a_{30}=0$$

$$2va_{22}= 3a_{03}+ a_{02}a_{10}+ a_{01}a_{11}+ a_{00}a_{12} \qquad (17)$$

$$6va_{32}= 2a_{10}a_{12}+ 2a_{01}a_{21}+ a_{11}^2+ 3a_{13}+ 2a_{02}a_{20}+ 2a_{00}a_{22}$$

$$- 3a_{10}a_{21}- 2a_{22}- 3a_{01}a_{30}- 3a_{00}a_{31}- 3a_{11}a_{20}=0$$

$$- a_{10}a_{22} - a_{11}a_{21} - a_{12}a_{20} - a_{23} - a_{02}a_{30}- a_{01}a_{31}- a_{00}a_{32}=0$$

$$6va_{33}= 2a_{10}a_{13}+ 2a_{02}a_{21}+ 2a_{11}a_{12}+ 2a_{03}a_{20}+ 2a_{01}a_{22}+ 2a_{00}a_{23}$$

$$- 2a_{20}^2- a_{31}=0$$

$$2va_{23}= a_{03}a_{10} + a_{02}a_{11}+a_{01}a_{12} + a_{00}a_{13}$$

solving the system of equations for the 12 unknown coefficients in function of $a_{00}$, $a_{01}$, $a_{10}$, $a_{11}$, and the $v$ parameter, result in the followings recurrence relation coefficients:

$$a_{02} =-(a_{00}^3 a_{10} + a_{01}a_{00}^2 + 4va_{00}a_{10}^2 + 2a_{11}va_{00} + 4a_{01}va_{10})/2v$$

$$a_{20}=(a_{01} + a_{00}a_{10})/2v$$

$$a_{21} =-(a_{00}^3 a_{10}+ a_{01}a_{00}^2 + 4va_{00}a_{10}^2 + va_{11}a_{00} + 3va_{01}a_{10})/2v^2$$

$$a_{12} =(a_{00}^4 a_{10} + a_{00}^3 a_{01}+ va_{00}^2 a_{10}^2 + va_{11}a_{00}^2 - 4va_{00}a_{01}a_{10}- 4va_{01}^2 - 2v^2 a_{11}a_{10})/(2v^2)$$

$$a_{22} =(3a_{00}^3 a_{10}^2 + 4a_{00}^2 a_{01}a_{10} + a_{00}a_{01}^2 + 6va_{00}a_{10}^3 + 4va_{01}a_{10}^2 - 2va_{11}a_{01})/(2v^2)$$

$$a_{30} =(a_{00}^2 a_{10} + a_{01}a_{00} + va_{10}^2 +v\, a_{11})/(6v^2)$$

$$a_{03} = (-a_{00}^5 a_{10} - a_{00}^4 a_{01} + 6va_{00}^3 a_{10}^2 - va_{11}a_{00}^3 + 13va_{00}^2 a_{01}a_{10} + 6va_{00}a_{01}^2 + 16v^2 a_{00}a_{10}^3 + 4v^2 a_{11}a_{00}a_{10} + 12 v^2 a_{01}a_{10}^2 - 6 v^2 a_{11}a_{01})/(6 v^2) \tag{18}$$

$$a_{31} = -(a_{00}^2 a_{10}^2 + 2a_{00}a_{01}a_{10} + a_{01}^2)/(2v^2)$$

$$a_{13} = (- 4a_{00}^5 a_{10}^2 - 6a_{00}^4 a_{01}a_{10} - 2a_{00}^3 a_{01}^2 - 16 va_{00}^3 a_{10}^3 + 5 va_{00}^3 a_{10}a_{11} + 7 va_{00}^2 a_{01}a_{11} + 29 va_{00}a_{01}^2 a_{10} - 24 v^2 a_{00}a_{10}^4 + 24 v^2 a_{00}a_{10}^2 a_{11} + 6 v^2 a_{00}a_{11}^2 + 15 va_{01}^3 - 16 v^2 a_{01}a_{10}^3 + 32 v^2 a_{01}a_{10} * a_{11})/(6a_{00} v^2)$$

$$a_{32} = (3a_{00}^5 a_{10}^2 - 3a_{00}^3 a_{01}^2 - 6va_{00}^3 a_{10}^3 + 5va_{00}^3 a_{10}a_{11} - 16va_{00}^2 a_{01}a_{10}^2 - va_{00}^2 a_{01}a_{11} + 11va_{00}a_{01}^2 a_{10} - 24v^2 a_{00}a_{10}^4 + 20 v^2 a_{00}a_{10}^2 a_{11} + 8 v^2 a_{00}a_{11}^2 + 15a_{01}^3 - 16 v^2 a_{01}a_{10}^3 + 32 v^2 a_{01}a_{10}a_{11})/(12a_{00} v^3)$$

$$a_{23} = (- 5a_{00}^5 a_{10}^2 - 4a_{00}^4 a_{01}a_{10} + a_{00}^3 a_{01}^2 - 10 va_{00}^3 a_{10}^3 + va_{11}a_{00}^3 a_{10} + 16 va_{00}^2 a_{01}a_{10}^2 + 7va_{11}a_{00}^2 a_{01} + 23 va_{00}a_{01}^2 a_{10} - 8 v^2 a_{00}a_{10}^4 + 16 v^2 a_{11}a_{00}a_{10}^2 + 3 va_{01}^3 - 4 v^2 a_{01}a^3 + 8 v^2 a_{11}a_{01}a_{10})/12 v^3$$

$$a_{33} = (- 3a_{00}^7 a_{10}^2 + 3a_{00}^5 a_{01}^2 + 12 va_{00}^5 a_{10}^3 + 15 va_{00}^5 a_{10}a_{11} + 86 va_{00}^4 a_{01}a_{10}^2 + 21 va_{00}^4 a_{01}a_{11} + 83 va_{00}^3 a_{01}^2 a_{10} + 24 v^2 a_{00}^3 a_{10}^4 + 72 v^2 a_{00}^3 a_{10}^2 a_{11} + 12 v^2 a_{00}^3 a_{11}^2 + 15 va_{00}^2 a_{01}^3 + 144v^2 a_{00}^2 a_{01}a_{10}^3 + 26 v^2 a_{00}^2 a_{01}a_{10}a_{11} + 130 v^2 a_{00}a_{01}^2 a_{10}^2 - 42 v^2 a_{00}a_{01}^2 a_{11} - 48 v^3 a_{00}a_{10}^5 + 48v^3 a_{00}a_{10}^3 a_{11} + 30 v^2 a_{01}^3 a_{10} - 32 v^3 a_{01}a_{10}^4 + 64 v^3 a_{01}a_{10}^2 a_{11})/(36a_{00} v^4)$$

we recover the coefficients of the stationary state as a particular case when we eliminates the coefficients of the *t* variable.

In the next example, we will try to solve the Korteweg–de Vries equation [25, 26]:

$$\frac{\partial U}{\partial t} + \frac{\partial^3 U}{\partial x^3} + 6U \frac{\partial U}{\partial x} = 0 \tag{19}$$

The equation admits a traveling wave solution. Then, we can to do the following transformation:

$$U(x,t) = \phi(z), \text{ where } z = kx - \lambda t \tag{20}$$

and $c = \lambda/k$ is the speed of light. Therefore, substituting Eq. (20) in Eq. (19) results the following non linear ODE:

$$-c\frac{d\phi}{dz} + k^2 \frac{d^3\phi}{dz^3} + 6\phi\frac{d\phi}{dz} = 0, \tag{21}$$

We solve it using the following proposal of PSS:

$$\phi(z) = \sum_{n=0}^{\infty} a_n z^n \qquad (22)$$

Substituting Eq. (22) in Eq. (21):

$$-c\sum_{n=1}^{\infty} n a_n z^{n-1} + k^2 \sum_{n=2}^{\infty} n(n-1)(n-2) a_n z^{n-3} + 6\sum_{n=0}^{\infty} a_n x^n \sum_{n=1}^{\infty} n a_n z^{n-1} = 0 \qquad (23)$$

Developing the series until reach $n=8$, and we obtain the following System of algebraic equations for the coefficients:

$$\begin{gathered}
\mathbf{n=0} \\
6a_0 a_1 + 6k^2 a_3 = c a_1 \\
\mathbf{n=1} \\
6a_1^2 + 24k^2 a_4 + 12 a_0 a_2 = 2 c a_2 \\
\mathbf{n=2} \\
60 k^2 a_5 + 18 a_0 a_3 + 18 a_1 a_2 = 3 c a_3 \\
\mathbf{n=3} \\
30 k^2 a_6 + 6 a_0 a_4 + 3 a_2^2 + 6 a_1 a_3 = c a_4 \\
\mathbf{n=4} \\
210 k^2 a_7 + 30 a_0 a_5 + 30 a_1 a_4 + 30 a_2 a_3 = 5 c a_5 \\
\mathbf{n=5} \\
18 a_3^2 + 336 k^2 a_8 + 36 a_0 a_6 + 36 a_1 a_5 + 36 a_2 a_4 = 6 c a_6 \\
\ldots
\end{gathered} \qquad (24)$$

The system of equations is solved considering the known parameters ($a_0$, $a_1$ and $a_2$) and the physical parameters (*k* and c):

$$a_3 = -(6 a_0 a_1 - c a_1)/6 k^2$$

$$a_4 = -(6 a_0 a_2 - c a_2 + 3 a_1^2)/12 k^2$$

$$a_5 = (36 a_0^2 a_1 - 12 c a_0 a_1 + c^2 a_1 - 36 k^2 a_1 a_2)/120 k^4$$

$$a_6 = (36 a_0^2 a_2 - 12 c a_0 a_2 + c^2 a_2 - 36 k^2 a_2^2 + 90 a_0 a_1^2 - 15 c a_1^2)/360 k^4 \qquad (25)$$

$$a_7 = (-216a_0{}^3 a_1 + 108 c a_0{}^2 a_1 - 18 c^2 a_0 a_1 - 15 c a_1{}^2 - 216 k^2 c a_1 a_2 + c^3 a_1 + 1296 k^2 a_0 a_1 a_2 + 180 k^2 a_1{}^3)/5040 k^6$$

$$a_8 = -(216 a_0{}^3 a_2 - 108 c a_0{}^2 a_2 + 18 c^2 a_0 a_2 + 63 c^2 a_1{}^2 - 216 k^2 c a_1 a_2 - c^3 a_2 + 1296 k^2 a_0 a_2{}^2 - 1188 k^2 a_1{}^2 a_2{}^2 + 2268 a_0{}^2 a_1{}^2 - 756 a_0 a_1{}^2)/20160 k^6$$

Therefore, our solution in function of the coefficients ($a_0$, $a_1$ and $a_2$) is:

$$U(x,t) = \phi(z) = a_0 + a_1(kx - wt) + a_2(kx - \lambda t)^2 + a_3(kx - \lambda t)^3 + a_4(kx - \lambda t)^4 + a_5(kx - \lambda t)^5 + ...$$

Starting from this solution, also we can to obtain a particular solution, just considering that the parameter $a_1=0$:

$$a_4 = -(6 a_0 a_2 - c a_2)/12 k^2$$

$$a_6 = (36 a_0{}^2 a_2 - 12 c a_0 a_2 - 36 a_2{}^2 k^2 + a_2 c^2)/360 k^4$$

$$a_8 = -(216 a_0{}^3 a_2 - 108 c a_0{}^2 a_2 - 1296 a_0 a_2{}^2 k^2 + 18 c^2 a_0 a_2 + 216 c a_2{}^2 k^2 - c^3 a_2)/20160 k^6$$

and the particular solution in function of the coefficients ($a_0$, $a_2$) is:

$$U(x,t) = \phi(z) = a_0 + a_2(kx - wt)^2 + a_4(kx - \lambda t)^4 + a_6(kx - \lambda t)^6 + a_8(kx - \lambda t)^8 + ...$$

The solution incorporates an even symmetry according with the parity property of the PDE in the Eq. (19).

In this last example, we solve the coupled Korteweg–de Vries equations [26, 27]:

$$\frac{\partial U}{\partial t} = \frac{1}{2}\frac{\partial^3 U}{\partial x^3} - 3U\frac{\partial U}{\partial x} + 3\frac{\partial(VW)}{\partial x},$$

$$\frac{\partial V}{\partial t} = -\frac{\partial^3 V}{\partial x^3} + 3U\frac{\partial V}{\partial x}, \qquad \frac{\partial W}{\partial t} = -\frac{\partial^3 W}{\partial x^3} + 3U\frac{\partial W}{\partial x}.$$

(26)

This system of equations also admits traveling wave solutions. Similarly, we can propose the transformation:

$$U(x,t) = u(z), \quad V(x,t) = v(z), \quad W(x,t) = w(z) \tag{27}$$

where $z = kx - \lambda t$, and $c = \lambda/k$ is the speed of the light. Substituting Eq. (27) in Eq. (26) results in the coupled system of ODE:

$$-c\frac{du}{dz} = \frac{k^2}{2}\frac{d^3u}{dz^3} - 3u\frac{du}{dz} + 3\frac{d(vw)}{dz}, \tag{28}$$

$$-c\frac{dv}{dz} = -k^2\frac{d^3v}{dz^3} + 3u\frac{dv}{dz}, \qquad -c\frac{dw}{dz} = -k^2\frac{d^3w}{dz^3} + 3u\frac{dw}{dz}.$$

Now we implement the use of a system of PSS functions:

$$u(z) = \sum_{n=0}^{\infty} a_n z^n, \quad v(z) = \sum_{n=0}^{\infty} b_n z^n \text{ and } \quad w(z) = \sum_{n=0}^{\infty} c_n z^n \tag{29}$$

and substituting Eq. (29) in (28), we obtain the following equations:

$$-c\sum_{n=1}^{\infty} na_n z^{n-1} = \frac{k^2}{2}\sum_{n=2}^{\infty} n(n-1)(n-2)a_n z^{n-3} - 3\sum_{n=0}^{\infty} a_n z^n \sum_{n=1}^{\infty} na_n z^{n-1}$$

$$+ 3\left(\sum_{n=1}^{\infty} nb_n z^{n-1}\sum_{n=0}^{\infty} c_n z^n + \sum_{n=0}^{\infty} b_n z^n \sum_{n=1}^{\infty} nc_n z^{n-1}\right)$$

$$-c\sum_{n=1}^{\infty} nb_n z^{n-1} = -k^2\sum_{n=2}^{\infty} n(n-1)(n-2)b_n z^{n-3} + 3\sum_{n=0}^{\infty} a_n z^n \sum_{n=1}^{\infty} nb_n z^{n-1} \tag{30}$$

$$-c\sum_{n=1}^{\infty} nc_n z^{n-1} = -k^2\sum_{n=2}^{\infty} n(n-1)(n-2)c_n z^{n-3} + 3\sum_{n=0}^{\infty} a_n z^n \sum_{n=1}^{\infty} nc_n z^{n-1}$$

Matching the coefficients of the same power degree until reach $n=8$, we obtain the following system of algebraic equations for the expansion series coefficients:

$$3a_0a_1 = ca_1 + 3b_0c_1 + 3b_1c_0 + 3a_3k_1^2$$

$$3a_1^2\ 6a_0a_2 = 12a_4k_1^2 + 2c\ a_2 + 6b_0c_2 + 6b_1c_1 + 6b_2c_0$$

$$9a_0a_3 + 9a_1a_2 = 30a_5k_1^2 + 3ca^3 + 9b_0c_3 + 9b_1c_2 + 9b_2c_1 + 9b_3c_0$$

$$6a_2^2 + 12a_0a_4 + 12a_1a_3 = 60k_1^2a_6 + 4ca_4 + 12b_0c_4 + 12b_1c_3 + 12b_2c_2 + 12b_3c_1 + 12b_4c_0$$

$15a_0a_5 + 15a_1a_4 + 15a_2a_3 = 105\ k_1^2 a_7 + 5c\ a_5 + 15b_0c_5 + 15b_1c_4 + 15b_2c_3 + 15b_3c_2 + 15b_4c_1 + 15b_5c_0$

$9a_3^2 + 18a_0a_6 + 18a_1a_5 + 18a_2a_4 = 168k_1^2 a_8 + 6\ ca_6 + 18b_0c_6 + 18b_1c_5 + 18b_2c_4 + 18b_3c_3 + 18b_4c_2 + 18b_5c_1 + 18b_6c_0$

$$6k_2^2 b_3 = cb_1 + 3a_0b_1$$

$$24k_2^2 b_4 = 6a_0b_2 + 3a_1b_1 + 2cb_2$$

$$60k_2^2 b_5 = 9a_0b_3 + 6a_1b_2 + 3a_2b_1 + 3cb_3 \qquad (31)$$

$$120\ k_2^2 b_6 = 12a_0b_4 + 9a_1b_3 + 6a_2b_2 + 3a_3b_1 + 4cb_4$$

$$210\ k_2^2 b_7 = 15a_0b_5 + 12a_1b_4 + 9a_2b_3 + 6a_3b_2 + 3a_4b_1 + 5cb_5$$

$$336\ k_2^2 b_8 = 18a_0b_6 + 15a_1b_5 + 12a_2b_4 + 9a_3b_3 + 6a_4b_2 + 3a_5b_1 + 6cb_6$$

$$6a_0c_1 + 6c_3k_3^2 = cc_1$$

$$24k_3^2 c_4 + 6a_1c_1 + 12a_0c_2 = 2cc_2$$

$$60k_3^2 c5 + 18a_0c_3 + 12a_1c_2 + 6a_2c_1 = 3cc_3$$

$$120k_3^2 c_6 + 24a_0c_4 + 18a_1c_3 + 12a_2c_2 + 6a_3c_1 = 4cc_4$$

$$210k_3^2 c_7 + 30a_0c_5 + 24a_1c_4 + 18a_2c_3 + 12a_3c_2 + 6a_4c_1 = 5cc_5$$

$$336\ k_3^2 c_8 + 36a_0c_6 + 30a_1c_5 + 24a_2c_4 + 18a_3c_3 + 12a_4c_2 + 6a_5c_1 = 6cc_6$$

resulting in 15 equations with a set of 15 unknowns coefficients. Therefore we can to solve them using Matlab® solver, obtaining the expansion coefficients in function of the known coefficients ($a_0$, $a_1$, $a_2$, , $b_0$, $b_1$, $b_2$, $c_0$, $c_1$, $c_2$) and their physical parameters ($c$, $k$):

$$a_3 = -(ca_1 - 3a_0a_1 + 3b_0c_1 + 3b_1c_0)/(3k_1^2)$$

$$a_4 = -(-3a_1^2 - 6a_0a_2 + 2ca_2 + 6b_0c_2 + 6b_1c_1 + 6b_2c_0)/(12k_1^2)$$

$a_5 = -(12ck_2^2 k_3^2 a_0a_1 - 2a_1c^2 k_2^2 * k_3^2 - 18a_0^2 a_1 k_2^2 k_3^2 - 18\ k_1^2 k_2^2 a_0b_0c_1 + 9\ k_1^2 k_3^2 a_0b_1c_0 + 18\ k_2^2 k_3^2 a_0b_0c_1 + 18a_0b_1c_0k_2^2 k_3^2 + 3c\ k_1^2 k_2^2 b_0c_1 + 3ck_1^2 k_3^2 b_1c_0 - 6c\ k_2^2 k_3^2 b_0c_1 - 6c\ b_1c_0k_2^2 k_3^2 - 18\ k_1^2 k_2^2 k_3^2 a_1a_2 + 18\ k_1^2 k_2^2 k_3^2 b_1c_2 + 18b_2c_1k_1^2 k_2^2 k_3^2)/(60k_1^4 k_2^2 k_3^2)$

$$a_6 = -(30\, ck_2^2 k_3^2 a_1^2 - 36\, k_2^2 k_3^2 a_0^2 a_2 - 90 a_0 a_1^2 k_2^2 k_3^2 - 4 a_2 c^2 k_2^2 k_3^2 - 36 a_2^2 k_1^2 k_2^2 k_3^2 +$$
$$24 a_0 a_2 c k_2^2 k_3^2 - 36 a_0 b_0 c_2 k_1^2 {*} k_2^2 - 72 a_0 b_1 c_1 k_1^2 k_2^2 - 18 a_1 b_0 c_1 k_1^2 k_2^2 + 36 a_0 b_1 c_1 k_1^2 k_3^2 +$$
$$18 a_0 b_2 c_0 k_1^2 k_3^2 + 9 a_1 b_1 c_0 k_1^2 k_3^2 + 36\, k_2^2 k_3^2 a_0 b_0 c_2 + 36\, k_2^2 k_3^2 a_0 b_1 c_1 + 36\, k_2^2 k_3^2 a_0 b_2 c_0 + 72$$
$$k_2^2 k_3^2 a_1 b_0 c_1 + 72\, k_2^2 k_3^2 a_1 b_1 c_0 + 6 c\, k_1^2 k_2^2 b_0 c_2 + 12 c\, k_1^2 k_2^2 b_1 c_1 + 12 c b_1 c_1 k_1^2 k_3^2 + 6 c k_1^2 k_3^2$$
$$b_2 c_0 - 12 c k_2^2 k_3^2 b_0 c_2 - 12 c k_2^2 k_3^2 b_1 c_1 - 12 c\, k_2^2 k_3^2 b_2 c_0 + 72\, k_1^2 k_2^2 k_3^2 b_2 c_2)/(360 k_1^4 k_2^2 k_3^2)$$

$$b_3 = (3 a_0 b_1 + c b_1)/(6 k_2^2) \tag{32}$$

$$b_4 = (6 a_0 b_2 + 3 a_1 b_1 + 2 c b_2)/(24 k_2^2)$$

$$b_5 = (9 a_0^2 b_1 + c^2 b_1 + 6 c a_0 b_1 + 12 a_1 b_2 k_2^2 + 6 a_2 b_1 k_2^2)/(120 k_2^4)$$

$$b_6 = (9\, k_1^2 b_2 a_0^2 + 18 a_1 a_0 b_1 k_1^2 + 9 k_2^2 a_1 a_0 b_1 + 6 c k_1^2 b_2 a_0 - 9\, k_2^2 c_0 b_1^2 + 6 c a_1 b_1 k_1^2 - 3 c k_2^2 a_1 b_1 -$$
$$9 k_2^2 b_0 c_1 b_1 + c^2 k_1^2 b_2 + 18 k_1^2 k_2^2 a_2 b_2)/(360 k_1^2 k_2^4)$$

$$c_3 = -(6 a_0 c_1 - c c_1)/(6 k_3^2)$$

$$c_4 = -(6 a_0 c_2 + 3 a_1 c_1 - c c_2)/(12 k_3^2)$$

$$c_5 = -(12 c a_0 c_1 - c^2 c_1 - 36 a_{02} c_1 + 24 k_3^2 a_1 c_2 + 12\, k_3^2 a_2 c_1)/(120 k_3^4)$$

$$c_6 = (36 c_2 a_0^2 k_1^2 - 12 c k_1^2 c_2 a_0 + 72 k_1^2 a_1 a_0 c_1 - 18\, k_3^2 a_1 a_0 c_1 + c^2 k_1^2 c_2 - 12 c k_1^2 a_1 c_1 + 6 c k_3^2 a_1 c_1 +$$
$$18 k_3^2 b_0 c_1^2 + 18 k_3^2 b_1 c_0 c_1 - 36 k_1^2 k_3^2 a_2 c_2)/(360 k_1^2 k_3^4).$$

Therefore, with these coefficients we can to obtain the following solutions:

$$U(x,t) = a_0 + a_1(kx - wt) + a_2(kx - \lambda t)^2 + a_3(kx - \lambda t)^3 + a_4(kx - \lambda t)^4 + a_5(kx - \lambda t)^5 + \ldots$$

$$V(x,t) = b_0 + b_1(kx - wt) + b_2(kx - \lambda t)^2 + b_3(kx - \lambda t)^3 + b_4(kx - \lambda t)^4 + b_5(kx - \lambda t)^5 + \ldots$$

$$W(x,t) = c_0 + c_1(kx - wt) + c_2(kx - \lambda t)^2 + c_3(kx - \lambda t)^3 + c_4(kx - \lambda t)^4 + c_5(kx - \lambda t)^5 + \ldots$$

In a similar way with the previous problems, we can to considerer the particular case ($a_1 = b_1 = c_1 = 0$), and to obtain an even solution for $U$, $V$ and $W$ because the symmetry of the system of equations is odd.

## 4 Discussion and conclusions

In this letter we have showed that is possible to solve non linear DE with the PSSM. This is implemented as a general approximate solution for each one, PDE, or ODE, in a similar way than the solution of Linear ODE. So, we transform one DE problem into an algebraic system of equations. Therefore, with this method, it is possible to obtain a well posed problem, because here, when the system of DE is closed, *i. e.* it have the same variable function than equations, then the number of algebraic variables (expansion series coefficients) that we obtain with this method, is the same with respect to the equations number. Thus, according with the methodology described, it is possible to obtain a solution.

This PSSM is a semi analytic technique that could permits to obtain, in an easier and exact way, the solution of difficult differential equations with an approximated closed form expression. This is especially useful to solve non linear equations, opening the possibility to describe in an exact and convergent way, the behavior of chaotic dynamical systems [14, 16].

Although it was not the focus of this article, the solution could be approximated until the degree necessary into the power series. The convergent of the PSSM depend, of the number of terms used of the series. Once we know it, we can to determine the domain of the space and time where the solution is valid.

In summary, we have shown that the PSSM is a general technique which can be used to solve any kind of non linear differential equations [17]. This opens the possibility of analyze other characteristics of the NLPDE, and obtaining a better semi analytic approximations that involves less computational efforts. This procedure will provide much greater accuracy just adding more terms into the series.

# 5 Acknowledgements

The authors are extremely grateful to Roman López Sandoval and Rogelio Ospina Ospina for their very careful reading and helpful discussion of the manuscript.
Also E. L. S. would like to acknowledge the support of the Brazilian agency Conselho Nacional de Desenvolvimento Científico e Tecnológico (CNPq) with the Grant of the PCI D-B from 01/01/2012 until now.